\documentclass[useAMS,usenatbib]{mn2e}
\usepackage{epsfig}

\title[Frequency of large variations]{The frequency of large variations 
in the near-infrared fluxes of T Tauri stars}

\author[Scholz]{Alexander Scholz$^{1}$\thanks{E-mail: aleks@cp.dias.ie}\\
$^{1}$ School of Cosmic Physics, Dublin Institute for Advanced Studies, 31 Fitzwilliam Place, 
Dublin 2, Ireland}

\begin{document}

\date{Accepted. Received.}

\pagerange{\pageref{firstpage}--\pageref{lastpage}} \pubyear{2002}

\maketitle

\label{firstpage}

\begin{abstract}
Variability is a characteristic feature of young stellar objects (YSOs) and could contribute
to the large scatter observed in HR diagrams for star forming regions. For typical YSOs, however,
the long-term effects of variability are poorly constrained. Here I use archived near-infrared photometry 
from 2MASS, UKIDSS, and DENIS to investigate the long-term variability of high-confidence members
of the four star forming regions $\rho$-Oph, ONC, IC348, and NGC1333. The total sample comprises more
than 600 objects, from which $\sim 320$ are considered to have a disk. The dataset covers timescales
up to 8\,yr. About half of the YSOs are variable on a 2$\sigma$ level, with median amplitudes of 5-20\%. 
The fraction of highly variable objects with amplitudes $>0.5$\,mag in at least two near-infrared bands 
is very low -- 2\% for the entire sample and 3\% for objects with disks. These sources with strong 
variability are mostly objects with disks and are prime targets for follow-up studies. A transition disk 
candidate in IC348 is found to have strong K-band variations, likely originating in the disk. The 
variability amplitudes are largest in NGC1333, presumably because it is the youngest sample of YSOs. 
The frequency of highly variable objects also increases with the time window of the observations (from weeks 
to years). These results have three implications: 1) When deriving luminosities for YSOs from near-infrared 
magnitudes, the typical error introduced by variability is in the range of 5-20\% percent and depends on disk 
fraction and possibly age. 2) Variability is a negligible contribution to the scatter in HR diagrams of star 
forming regions (except for a small number of extreme objects), if luminosities are derived from near-infrared
magnitudes. 3) Accretion outbursts with an increase in mass accretion rate by several order of magnitudes, 
as required in scenarios for episodic accretion, occur with a duty cycle of $>2000-2500$\,yr in the Class 
II phase.
\end{abstract}


\section{Introduction}
\label{intro}

Before T Tauri stars were recognised as young stellar objects (YSOs) they were known primarily for their 
characteristic photometric variability. T Tau itself, first seen by \citet{1852AN.....35..371H}, was 
described in a paper by \citet{1891Obs....14...97K} as a 'remarkable variable star' with 'a magnitude 
range of from 9.4 to 13.5 or 14' \citep[see also the discussion in][]{1895MNRAS..55..442B}. About fifty years 
later, \citet{1945ApJ...102..168J} defined the class of T Tauri variables based on this prototype and a number 
of objects with similar lightcurves, including RW Aur, R CrA, RU Lup, RY Tau, UX Tau -- today some of the 
best-studied YSOs. They exhibit large-scale ($>1$ mag), irregular variations in visual bands on 
timescales ranging from hours to years. This type of variability can be attributed to the 
process of accretion from the circumstellar disk onto the newly formed star 
\citep[e.g.][]{1989A&ARv...1..291A,1994AJ....108.1906H,1995A&A...299...89B}.

Nowadays, most of the young stellar objects with disks are identified by other means, primarily from 
their mid-infrared colour excess due to hot dust in the inner disk or from optical/near-infrared 
emission lines originating in the accretion flow or in the wind. It is well-established that a
large fraction of YSOs show variability on timescales of days to weeks with lightcurve
RMS of typically $<0.2$\,mag \citep[e.g.][]{2004A&A...417..557L}. In many cases this moderate variability 
is periodic and can be explained by cool and/or hot spots co-rotating with the objects
\citep{2007prpl.conf..297H}. In addition, it is known that at least some T Tauri stars exhibit stable 
long-term variations with large amplitudes of up to several magnitudes \citep[e.g.][]{2007A&A...461..183G}. More
recently, it has been established that these two types of variability occur over a wide range of
stellar masses, including brown dwarfs with $M<0.08\,M_{\odot}$ \citep{2004A&A...419..249S,2005A&A...429.1007S}.

For timescales exceeding a few weeks, however, there are little constraints on the ubiquity of T Tauri-type 
variability in {\it typical} YSOs. In this paper I analyse the near-infrared photometry from 2MASS, 
UKIDSS, and DENIS for large samples of T Tauri stars, to put limits on the evolution of their brightness 
on timescales of 2-8\,yr. In Sect. \ref{s1} I discuss the data used for this study. In the following 
Section \ref{s2} I select variable stars for YSOs in four different star forming regions. These results 
have implications for the interpretation of HR diagrams of these regions as well as for our understanding 
of the time evolution of accretion (see Sect. \ref{s3}).

\section{Data}
\label{s1}

The photometric data used for this study comes from the three projects 2MASS, UKIDSS, and DENIS. 
2MASS\footnote{2MASS is a joint project of the University of 
Massachusetts and the Infrared Processing and Analysis Center/California Institute of Technology, funded by 
the National Aeronautics and Space Administration and the National Science Foundation.}
is an all-sky survey carried out from 1997 to 2001 in the J-, H- and K-bands \citep{2006AJ....131.1163S}.
The 10$\sigma$ depth is approximately 16, 15, and 14\,mag in these three filters. 

UKIDSS is a seven-year survey program started in 2005 \citep{2007MNRAS.379.1599L}, carried out with the 
UKIRT Wide-Field Camera \citep{2007A&A...467..777C} in the five bands Y, Z, J, H, and K \citep{2006MNRAS.367..454H}. 
The project comprises five different survey components; for this paper we use the 6th Data Release of the 
Galactic Cluster Survey (for $\rho$-Oph, ONC, IC348) and the Galactic Plane Survey \citep[for NGC1333,][]{2008MNRAS.391..136L}
For more information on UKIDSS see for example \citet{2006MNRAS.372.1227D} and \citet{2008MNRAS.384..637H}.

DENIS is a survey of the southern sky in two near-infrared bands (J and K) and one optical band (Gunn-I), carried
out betwen 1996 and 2001 \citep{1999A&A...349..236E}. For this paper we use the 3rd data release from September 2005.

The near-infrared bands employed by the three surveys are similar, but not identical. I convert the DENIS
magnitudes to the 2MASS system using the transformations given by \citet{2001AJ....121.2851C}. The UKIDSS photometry 
was shifted into the 2MASS system using the conversion equations published by \citet{2006MNRAS.367..454H}. Many YSOs in
highly extincted regions have near-infrared colours for which these transformations are not properly calibrated;
this could lead to colour-dependent offsets between the two photometric systems. In some of the figures (see in
particular Fig. \ref{f1} and \ref{f2}), the cumulation of objects around the origin appears elongated, which could
be caused by such systematic errors.

UKIDSS is deeper than 2MASS and DENIS, but also affected by saturation at the bright end. The effects of
saturation are clearly visible when plotting magnitude differences between 2MASS and UKIDSS; above a certain 
limit UKIDSS underestimates the fluxes of the stars. When identifying variable sources, a magnitude 
cutoff was implemented to avoid bright stars for which saturation plays a role. 

For all archives no additional
cuts on signal-to-noise ratio were implemented, i.e. all objects with valid flux and error measurements were 
used for the analysis (e.g., the 2MASS quality flags A-E). The objects that turn out to be highly variable
and are listed in Table all have 2MASS photometry flags A or B -- signal-to-noise ratio $\ge 7$ -- in the
analysed bands.

The magnitude differences between the various surveys comprise a) variability, b) photometric uncertainties, and
c) systematic effects (saturation, band offsets). The photometric uncertainties in 2MASS and DENIS range
from a few percent up to 0.2\,mag for the objects investigated here; for UKIDSS these numbers are significantly
smaller. Systematic effects can contribute up to 0.1-0.2\,mag. Thus, any magnitude difference $>0.5$\,mag can safely
be attributed to a variable source. In this paper, I focus mostly on these strongly variable objects which can
unambiguously be identified from the 2-3 epochs available in the surveys. To exclude spurious detections, I 
require that an object shows this level of variation in at least two bands.

As the main diagnostic to identify variables and explore possible causes of their variability, I plot
the magnitude difference between two surveys in the K-band vs. the colour difference (either in $H-K$ or 
in $J-K$). These diagrams (see Fig. \ref{f1}, \ref{f3}, \ref{f5}, \ref{f6}) show the trends in the evolution
of the brightness (brighter/fainter) and colour (redder/bluer). Depending on the origin of the variability,
objects will appear in different quadrants: Hot spots due to accretion will make the objects brighter and bluer 
(upper right quadrant) or fainter and redder (lower left quadrant). Cool spots due to magnetic activity 
will have the same, albeit much smaller effect. Variable extinction, for example due to inhomogenities in
the disk, would again make the sources either brighter and bluer or fainter and redder. In contrast, variable
emission from the disk would cause variability towards brighter and redder (upper left quadrant) or fainter 
and bluer (lower right quadrant). Whatever the origin of the variations, it is expected that the sign of the
magnitude change is the same in all bands, which serves as an additional sanity check. For a more detailed 
discussion of the variability causes and their effects on magnitudes and colours see 
\citet{2001AJ....121.3160C,2009MNRAS.398..873S}.

To identify the objects with disks in the samples in $\rho$-Oph and IC348, I made use of the 'HREL' photometry 
catalogues from the 'Cores to Disks' (C2D) Spitzer Legacy Program \citep{2009ApJS..181..321E}.

\section{Identifying variable stars}
\label{s2}

\begin{table*}
\caption{Sources with strong variability in the regions $\rho$-Oph, ONC, IC348, NGC1333. Most were found
by comparing 2MASS and UKIDSS photometry (time window 6-8\,yr); the two exceptions come from the DENIS vs. 
2MASS comparison and are marked. 
\label{var}}
\begin{tabular}{llcccccll}
\hline
$\alpha$ (J2000) & $\delta$ (J2000) & $K2M$ (mag) & $dK$ (mag) & $dH$ (mag) & $dJ$ (mag) & Disk & SIMBAD names & Comments\\
\hline
16 26 17.23 & -24 23 45.4 & 12.251 & +0.515 & +0.882 &        & Y & ISO-Oph-21	        & faint in 2MASS\\
16 26 44.19 & -24 34 48.3 & 11.603 & -1.037 & -1.510 &        & Y & ISO-Oph-65, [GY92 ]111  &		\\
16 27 11.18 & -24 40 46.7 & 10.196 & -0.786 & -0.804 &        & Y & ISO-Oph-112, [GY92] 224 &		\\
16 27 17.59 & -24 05 13.7 & 10.727 & +0.910 & +0.695 & +0.632 & Y & ISO-Oph-123	        &		\\
16 27 26.94 & -24 40 50.8 &  9.745 & -0.650 & -0.713 &        & Y & ISO-Oph-141, [GY92] 265 & companion     \\
\hline
05 35 18.78 & -05 18 02.6 & 10.974 & -0.635 &        & -0.546 & ? & COUP 990, [H97b] 5058   &               \\
05 35 29.46 & -05 18 45.8 & 10.825 & +0.639 &        & +1.235 & ? & COUP 1425, [H97b] 3042  &               \\
05 34 51.82 & -05 21 39.0 & 12.411 & +0.688 &        & +0.674 & N? & COUP 64, [H97b] 124     & M3.5, DENIS   \\
05 35 05.38 & -05 24 10.5 &  9.616 & -0.505 &        & -0.687 & Y & COUP 236, [H97b] 265    & M0, DENIS	\\
\hline 
03 44 31.36 & +32 00 14.7 & 10.636 & -0.784 & -0.771 & -0.752 & Y & Cl* IC 348 LRL 55       & M0.5            \\
\hline
03 28 58.42 & +31 22 17.7 & 11.849 & -1.227 & -0.737 &        & Y & [LAL96] 166, MBO 38     & Class I       \\
03 29 03.13 & +31 22 38.2 & 11.323 & -0.620 & -0.828 & -0.581 & Y & [LAL96] 189, MBO 31     & Class II      \\
03 29 20.05 & +31 24 07.6 & 12.042 & -1.555 & -2.128 &        & Y & [LAL96] 296, MBO 46     & Class II      \\
\hline
\end{tabular}
\end{table*}

\subsection{$\rho$-Ophiuchus}

$\rho$-Ophiuchus (short $\rho$-Oph) is a region of star forming activity at a distance of $\sim 130$\,pc. The 
area is visible from the northern and southern hemisphere and was covered by 2MASS, DENIS, and UKIDSS/GCS. According 
to the census given in \citet{2008hsf2.book..351W}, the main cloud of $\rho$-Oph, L1688, harbours at least 300 YSOs, 
confirmed by their X-ray emission, infrared excess emission, Li absorption, H$\alpha$ emission or other signs 
of youth. Due to its high extinction, the cloud effectively acts as a screen, i.e. the sample should be clean
from background contamination. The median age of YSOs in L1688 is often cited as $<1$\,Myr, but a fraction
of the YSOs in this area might actually belong to an older population associated with the UpSco star forming
region (age 2-5\,Myr). For more information, see \citet{2008hsf2.book..351W}.

I start with the list of 316 members with 2MASS counterparts from \citet{2008hsf2.book..351W}. For the
comparison between UKIDSS and 2MASS I select the 223 objects with photometry in the H- and K-band 
in the two surveys. In the J-band, many objects are undetected in 2MASS due to the high extinction. In the 
UKIDSS dataset, all objects with $H<11$ and $K<9$ are affected by saturation; this brings the number of 
usable objects down to 144. 

\begin{figure*}
\includegraphics[width=6.1cm,angle=-90]{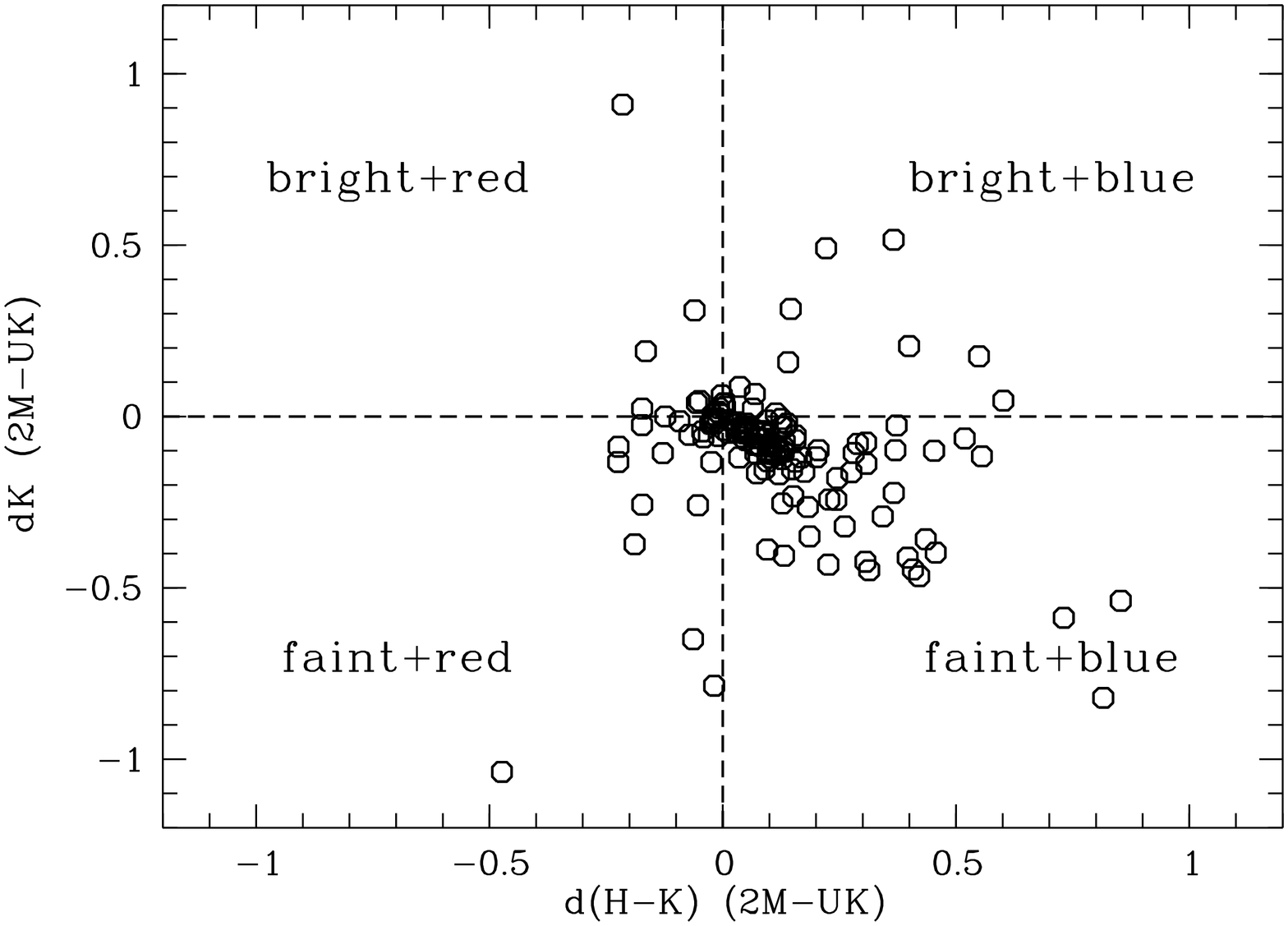} \hfill
\includegraphics[width=6.1cm,angle=-90]{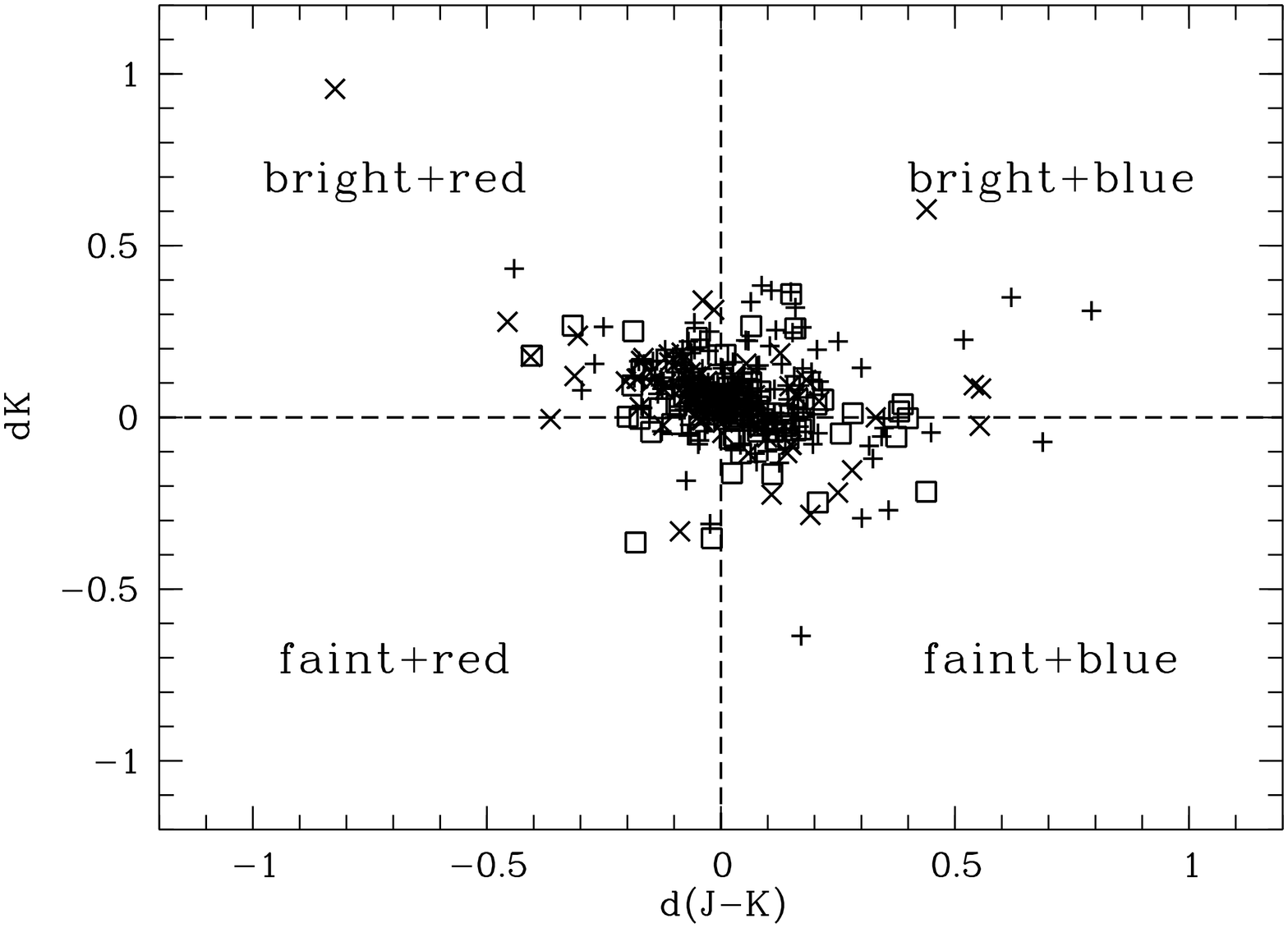} 
\caption{K-band variability vs. colour variability for YSOs in $\rho$\,Oph. The left panel shows variations on
timescales of $\sim 6$\,yr; the right panel shorter timescales of $<2$\,yr. Different symbols are used for
various combinations of datasets: circles - 2MASS vs. UKIDSS; crosses - 2MASS vs. DENIS epoch 1; plusses -
2MASS vs. DENIS epoch 2; squares - DENIS epoch 1 vs. epoch 2. 
\label{f1}}
\end{figure*}

In Fig. \ref{f1}, left panel, I show the K-band variability $dK$ vs. the colour variability $d(H-K)$for 
the UKIDSS/2MASS comparison. In this plot, objects in the upper/lower right/left quadrant became bright/faint 
and blue/red, respectively. The average epoch difference between 2MASS and UKIDSS is $\sim 6$\,yr.
8 objects in this sample show K-band variations of $dK>0.5$\,mag, two of them have $dK \sim 1$\,mag.
The maximum amplitude in the H-band is $dH = 1.5$\,mag. Five out of 8 have $>0.5$\,mag variations
in at least two bands and are listed in Table \ref{var}.

In the sample of 144 objects shown in Fig. \ref{f1}, left panel, 80 have Spitzer/IRAC colour excess 
([5.8\,$\mu m$]-[8.0\,$\mu m$]\,$>0.3$\,mag, [3.6\,$\mu m$]-[4.5\,$\mu m$]\,$>0.2$\,mag), interpreted as evidence
for a disk, among them all 5 highly variable sources.

$\rho$-Oph was covered twice by DENIS, with epoch differences relative to 2MASS of $-1$ to $+2$\,yr. 
From the initial sample of 316 members, 82 have usable photometry in 2MASS and DENIS, epoch 1,
194 in 2MASS and DENIS, epoch 2, and 78 in the two epochs of DENIS. In Fig. \ref{f1}, right panel, I
show the $dK$ vs. $d(J-K)$ plot for the three possible combinations from these datasets, covering timescales
$<2$\,yr. Three objects exhibit magnitude changes of $dK>0.5$\,mag, two of them with disks, but only
one in more than one band. This one, however, is at the detection limit of 2MASS in the J-band with 
signal-to-noise ratio of only 4. Therefore we do not list any new objects from this search in Table 
\ref{var}.

The 5 highly variable sources from the UKIDSS vs. 2MASS comparison are listed in Table \ref{var} and were
verified in the survey images. One of them is close to the sensitivity limit of 2MASS, although the 
photometric uncertainties are still well below 0.5\,mag. One more has a companion which is not well-resolved
in the 2MASS images, hence, it appears fainter in UKIDSS with its better resolution. Thus, the total fraction 
of sources in $\rho$-Oph that are highly variable on timescales of $\sim 6$\,yr is 3-5 out of 144 in the total 
sample (2-3\%), and 3-5 out of 80 with disks (4-6\%).

1 of the 5 $\rho$-Oph objects listed in Table \ref{var}, ISO-Oph-112, has already been identified by 
\citet{2008A&A...485..155A} as highly variable on timescales of up to 1\,yr (for reference, the no. in
their Table 4 is 58). Similar to the results presented here, \citet{2008A&A...485..155A} find relatively 
few objects with H- or K-band amplitudes exceeding 0.5\,mag.

\begin{figure}
\includegraphics[width=6.1cm,angle=-90]{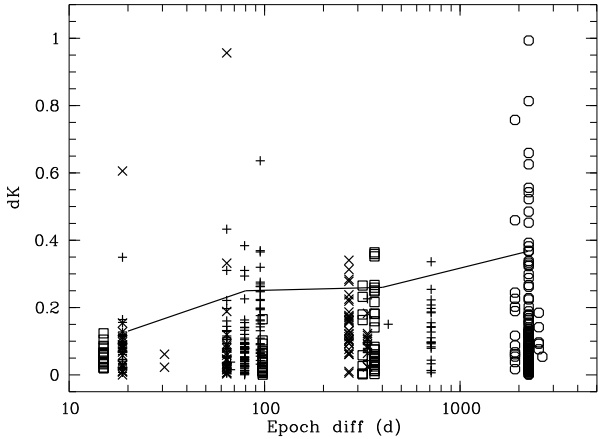} 
\caption{K-band magnitude change as a function of timescale for YSOs in $\rho$\,Oph. The plot uses data from 
2MASS, DENIS, and UKIDSS. Symbols are the same as in Fig. \ref{f1}. The solid line shows the 90th percentile
for the epoch bins $<40$\,d, 60-100\,h, 200-800\,d, and 1000-3000\,d.
\label{f2}}
\end{figure}

In Fig. \ref{f2} the absolute K-band variability is plotted against the epoch difference, combining the data 
from all three surveys. With this figure I aim to test whether the variations depend on the timescale or not.
The fraction of objects with $dK >0.5$ is $0.8 \pm ^{0.3}_{0.5}$ for epoch differences $<1000$\,d and
$5.6 \pm ^{2.5}_{1.9}$ for epoch differences $>1000$\,d (errors are 1$\sigma$ binomial confidence intervals). 
I also overplot the 90th percentile for four bins in epoch difference: $<40$\,d,
60-100\,h, 200-800\,d, and 1000-3000\,d. These values increase from 0.13 on the shortest timescales
to 0.37 on the longest. This result is robust within the statistical uncertainties of these percentiles.

I conclude that high-level variations in the near-infrared are more frequent on timescales of years
than on timescales of weeks to months. 

\subsection{ONC}

The ONC is a massive cluster of YSOs in the Orion star forming region with an age of about 1\,Myr. The cluster 
has been observed by 2MASS in 2000, by UKIDSS/GCS in 2005-2006, and by DENIS between 1996 and early 1999, 
although the coverage by UKIDSS and DENIS is incomplete. A good starting point for a cluster census is the list 
of point sources in the X-ray catalogue from the COUP project \citep{2005ApJS..160..319G}. Among these $>1600$ 
objects are 918 with optical and near-infrared counterpart. As argued by \citet{2005ApJS..160..353G} 
these can be assumed to be safe members of the cluster. The sample is dominated by K-M stars, but
also includes some more massive objects.

\begin{figure*}
\includegraphics[width=6.1cm,angle=-90]{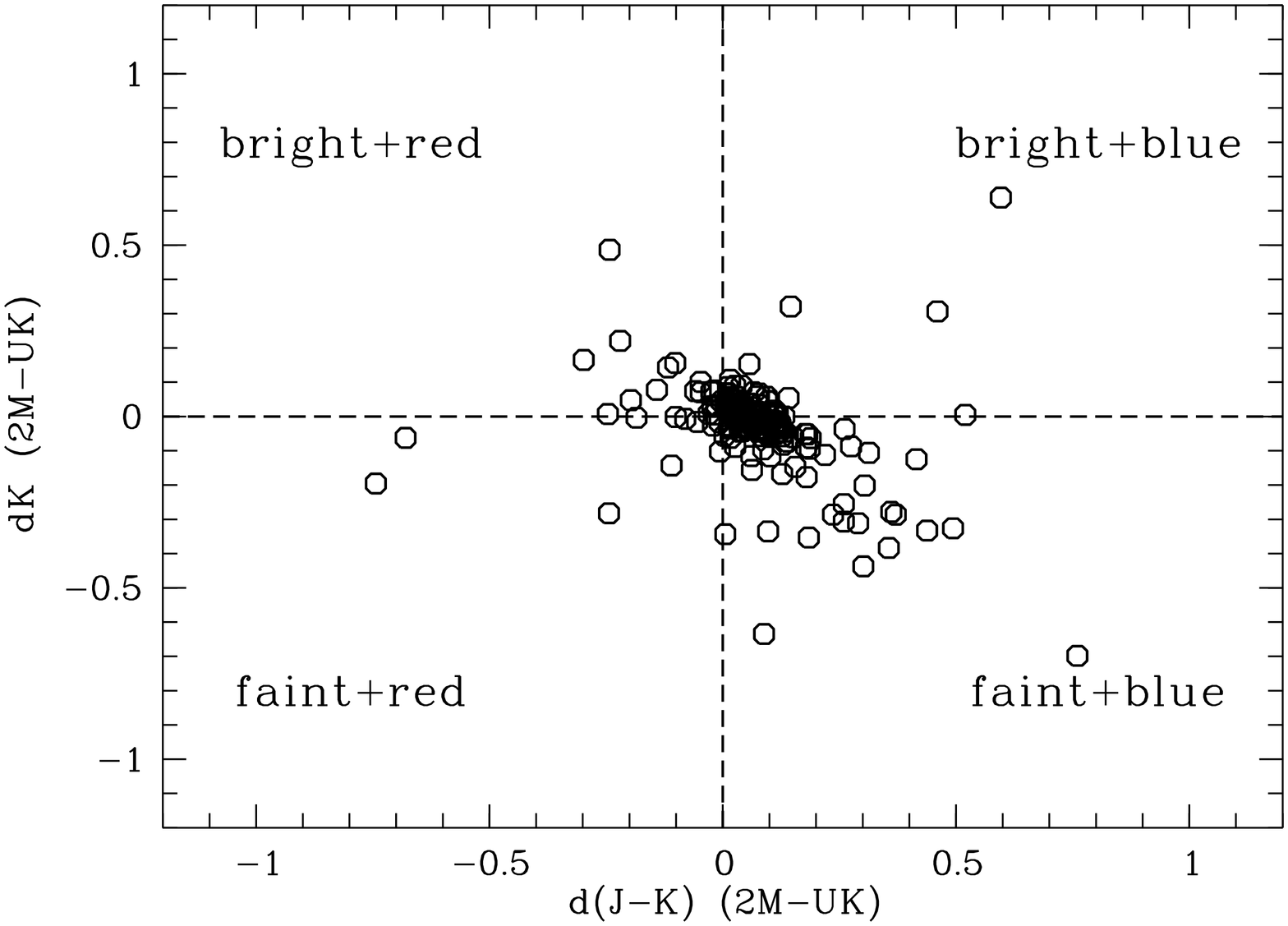}
\includegraphics[width=6.1cm,angle=-90]{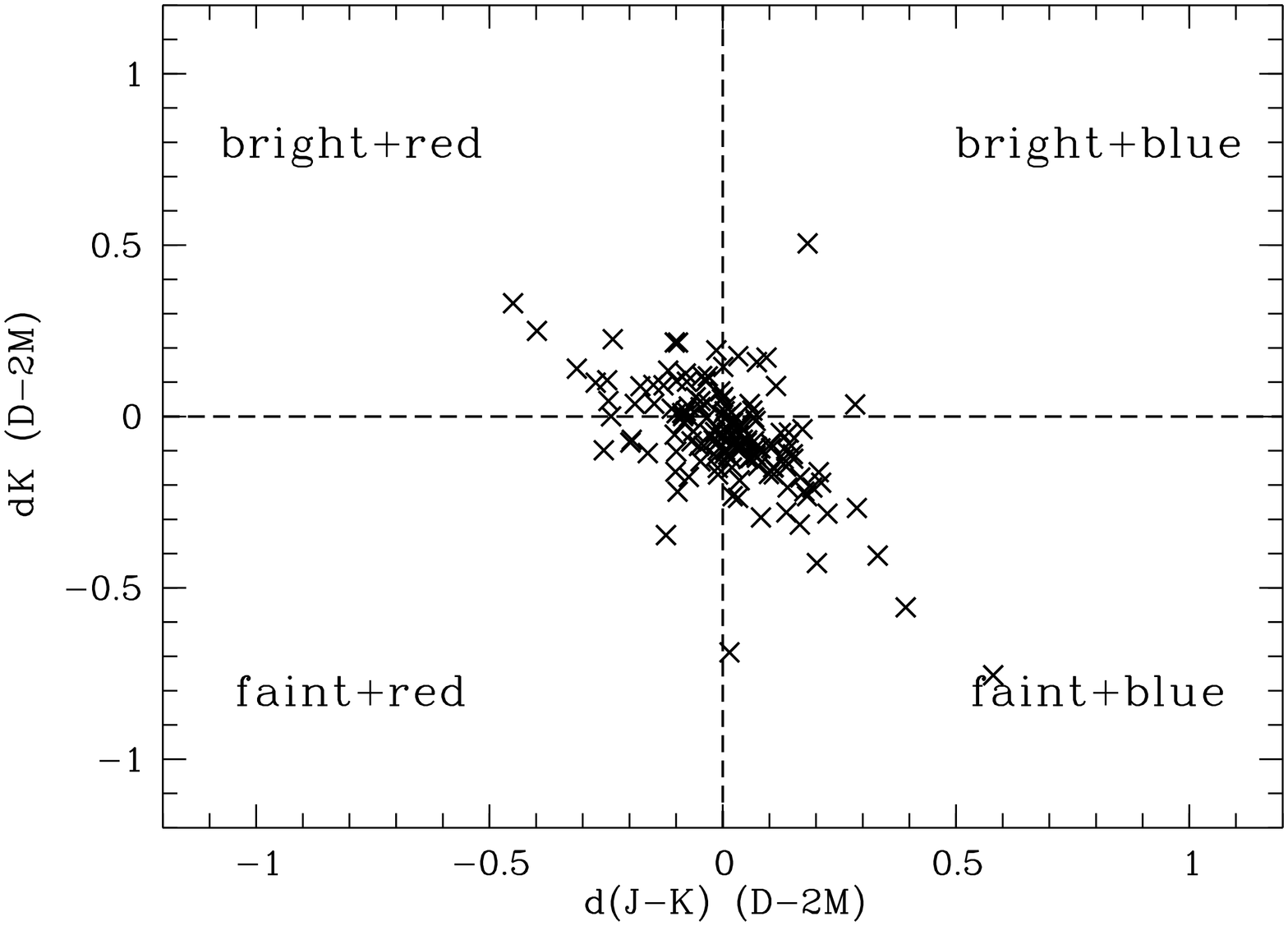}
\caption{K-band variability vs. colour variability for YSOs in the ONC. The left panel shows variations on
timescales of $\sim 6$\,yr; the right panel shorter timescales of $2-5$\,yr. Different symbols are used for
various combinations of datasets: circles - 2MASS vs. UKIDSS; crosses - DENIS vs. 2MASS.
\label{f3}}
\end{figure*}

Due to the limited coverage of UKIDSS, only 198 have J- and K-band photometry in 2MASS and UKIDSS,
from which 188 are unaffected by saturation ($J>11$, $K>9$). With H and K the numbers are even lower. 
As can be seen in Fig. \ref{f3}, left panel, three sources are strongly variable with $dK>0.5$\,mag. 
Two of them show this level of variability in two bands; they are listed in Table \ref{var}.

Note that no band transformation was carried out for the UKIDSS photometry in Fig. \ref{f3},
because not all objects in this plot have the required colour information. This causes some additional
scatter around the zeropoint in the diagram, but does not affect the selection of strongly variable
objects.

151 objects have usable J- and K-band photometry in 2MASS and DENIS, which is plotted in Fig. \ref{f3}, 
right panel. 4 of them exceed $dK=0.5$\,mag, 2 of them have also $dJ>0.5$ and are included in Table \ref{var}.

The ONC is a crowded area, and all of the objects listed in Table \ref{var} might be affected by close
neighbours, particularly in the 2MASS and DENIS images. Thus, some of these objects might not be variable.
Re-observations with good resolution are necessary to clarify their nature.

For the ONC we cannot probe the mid-infrared excess for the sample. The Spitzer data for this region
has not been published yet, and the region is not covered by the preliminary database from
the Wide-field Infrared Survey Explorer (WISE). Instead, we use the near-infrared excess $\Delta (I-K)$ from 
\citet{1998AJ....116.1816H} to identify the disks. In the sample plotted in Fig. \ref{f3}, left panel, 82 
objects have a measured $\Delta (I-K)$, for 40 of them it is $>0.3$\,mag, considered to be a safe disk 
detection by \citet{1998AJ....116.1816H}. For the right panel, 84 have a measurement, 44 of them $>0.3$\,mag.
Thus, the disk fraction in the samples plotted in Fig. \ref{f3} is in the range of 50\%.

For two of the four variables from Table \ref{var} the near-infrared excess is available in 
\citet{1998AJ....116.1816H}. One of them clearly has a disk (COUP 236), the other not (COUP 64). 
Lack of near-infrared excess does not necessarily imply the lack of circumstellar material, i.e. COUP 64 
might still harbour a disk. Three of the four objects listed in Table \ref{var} -- COUP 1425, 64, 236 -- 
are also identified as variables by \citet{2001AJ....121.3160C}.

In summary, the total fraction of objects with strong near-infrared variability on timescales of several years  
for the ONC is in the range of 1\% for all objects and 2\% for objects with disks.

\subsection{IC348}

For IC348, a $\sim 3$\,Myr old cluster in the Perseus star forming region, I start with the list of 288
spectroscopically confirmed cluster members published by \citet{2003ApJ...593.1093L}. Most of these objects 
are low-mass stars with spectral types K to M. The cluster was covered by 2MASS from late 1998 to early 1999 
and by UKIDSS/GCS in late 2006, i.e. the typical epoch difference is $\sim 8$\,yr.

\begin{figure*}
\includegraphics[width=6.1cm,angle=-90]{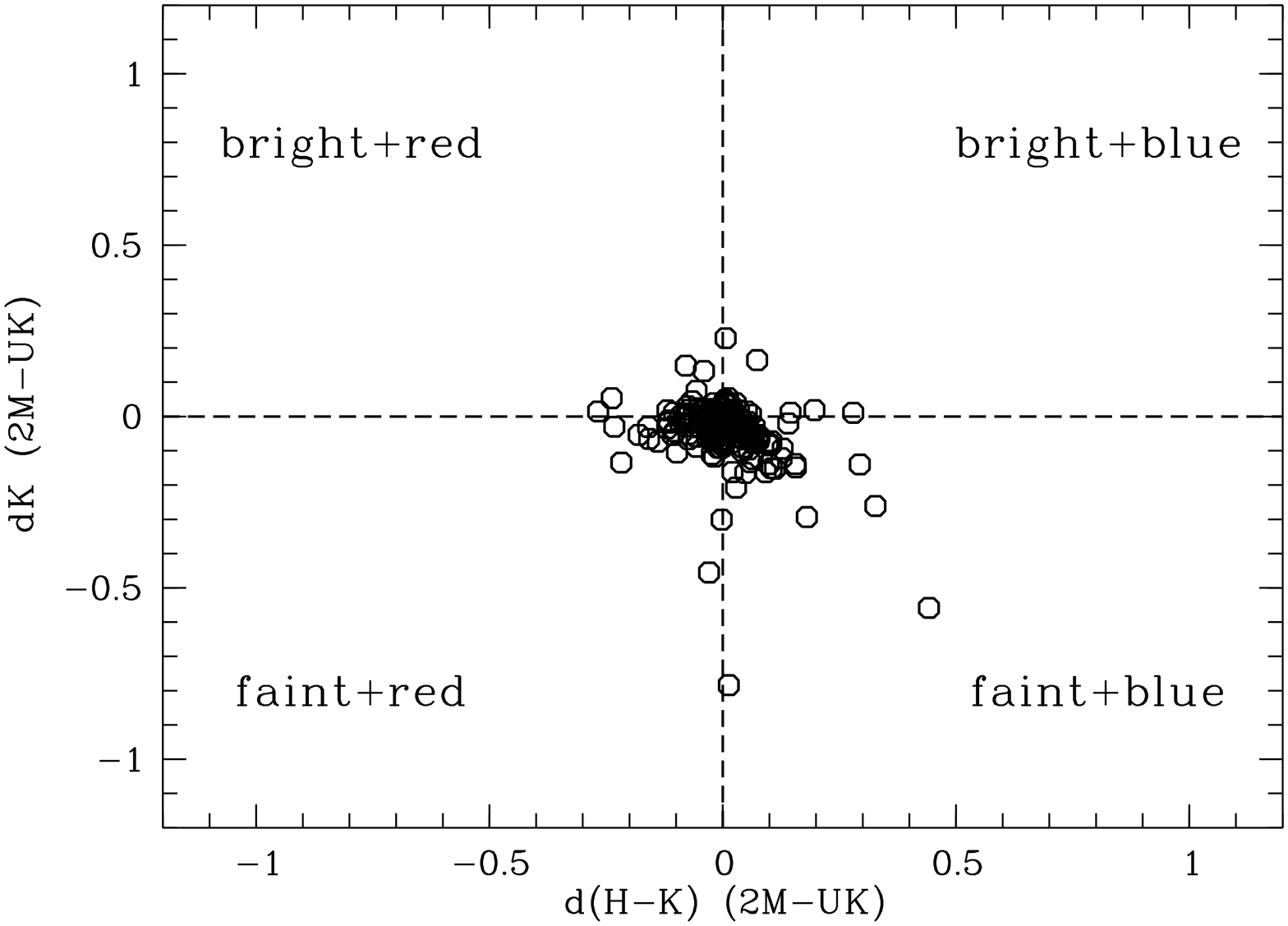} \hfill
\includegraphics[width=6.1cm,angle=-90]{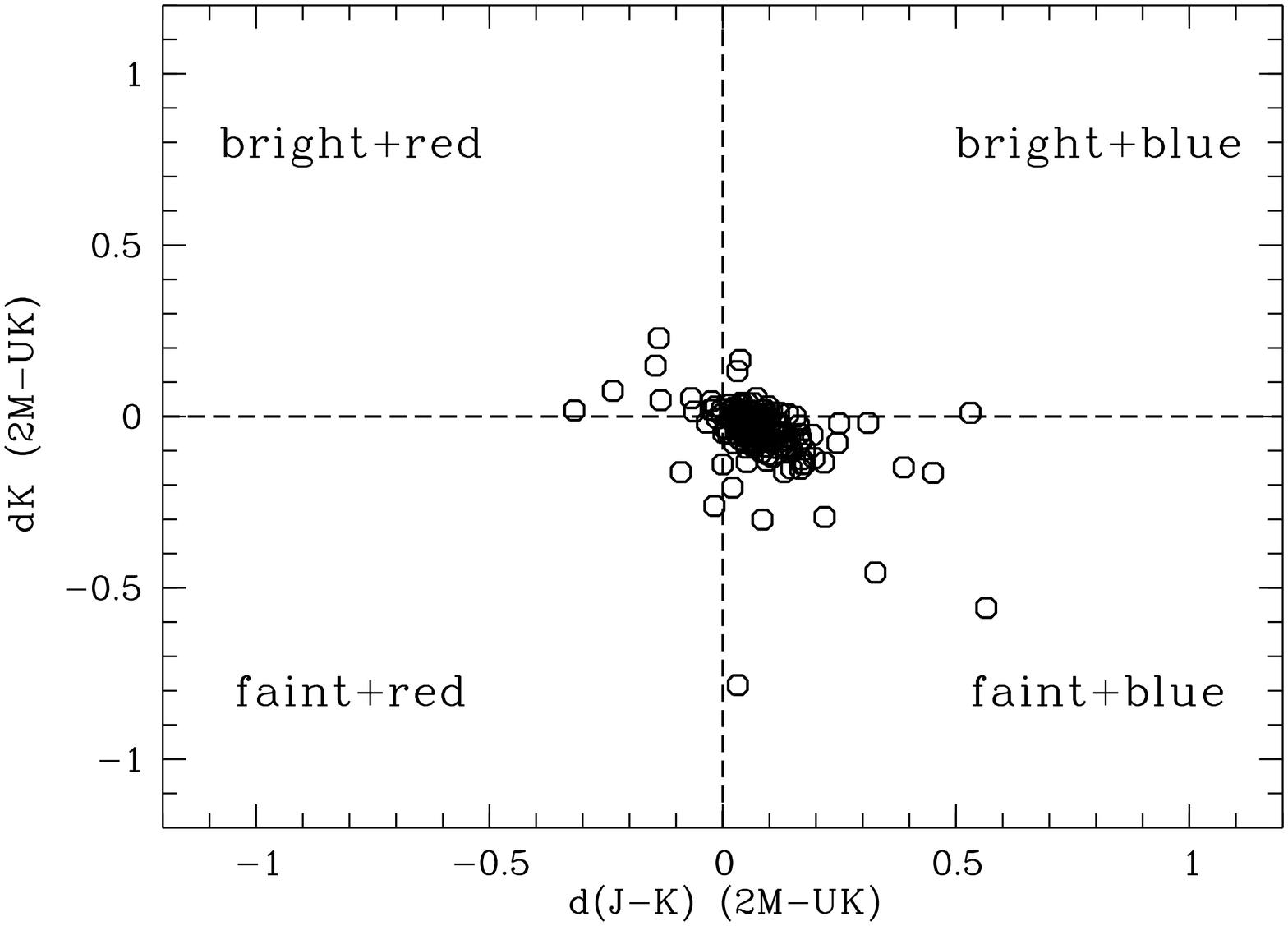} 
\caption{K-band variability vs. colour variability in $H-K$ (left panel) and $J-K$ (right panel) for YSOs 
in IC348 on timescales of $\sim 8$\,yr. The plots use data from 2MASS and UKIDSS.
\label{f5}}
\end{figure*}

249 objects have photometry in the H- and K-band in the two surveys, 201 of them are not affected 
by saturation in UKIDSS ($H>11$ and $K>10.5$). In Fig. \ref{f5}, left panel, we plot the K-band vs.
$H-K$ colour variability. 2 objects show K-band variations $>0.5$\,mag, one of them with $dK=0.8$ and
$dH=0.8$\,mag. This object is included in Table \ref{var}. 

The right panel in Fig. \ref{f5} shows the same plot using $J-K$ instead of $H-K$. Here 247 objects have
photometry in the two surveys, 208 without effects of saturation ($J>11$, $K>10.5$). Again, 
there are 2 objects with $dK>0.5$\,mag, the same ones as in the left panel.

From the samples shown in the figure, 63 have mid-infrared excess due to the presence of disks based
on the IRAC colours ([5.8\,$\mu m$]-[8.0\,$\mu m$]\,$>0.45$\,mag, [3.6\,$\mu m$]-[4.5\,$\mu m$]\,$>0.15$\,mag). 
Among them is 1 of the 2 highly variable objects, the one listed in Table \ref{var}. The second one
($\alpha = 03^h44^m22.57^s$, $\delta = +32^o01'53.7''$, SIMBAD name Cl* IC 348 LRL 72), has 
[5.8\,$\mu m$]-[8.0\,$\mu m$]\,$=0.44$\,mag and [3.6\,$\mu m$]-[4.5\,$\mu m$]\,$=0.01$\,mag. It also has strong 
excess at 24$\,\mu m$ with [8.0\,$\mu m$]-[24\,$\mu m$]\,$=4.8$\,mag, which is more than most object with disks 
in IC348. Thus, this source clearly has circumstellar material, but it might have an inner hole in the disk and 
therefore little excess at 3-8$\,\mu m$, a so-called transition disk. Indeed, the object was previously classified 
as a 'transition disk candidate' \citep{2010ApJ...708.1107M} and is an interesting target for further
monitoring. Since the variability for this source is mostly seen in the K-band, it is likely to originate 
in the disk.

The fraction of highly variable sources, as defined in this paper, it thus 1/208 for the total sample 
(1\%) and 1/63 for objects with disks (2\%).

In IC348 the plots indicate that there are little problems with the band correction, most likely because the 
extinction is on average lower than in $\rho$-Oph and the ONC. Therefore it is feasible to provide quantitative 
limits on the variability of the general population. The median difference between 2MASS and UKIDSS photometry
is 0.05, 0.04, 0.04\,mag in J, H, and K for the entire sample and 0.07, 0.06, 0.05 for the sources with disks.

These numbers are still affected by the observational errors, which are dominated by the 2MASS uncertainties.
The UKIDSS errors are typically by a factor of 10 lower than the 2MASS errors. The fractions of objects
for which the offsets are more then twice as large as the 2MASS errors (i.e. variable on a 2$\sigma$ level)
in J, H, K are 46\%, 41\%, 37\% for all objects and 56\%, 52\%, 60\% for the ones with disks. The
typical $1\sigma$ binomial uncertainties in these fractions are 4\% for all objects and 6\% for the 
stars with disks. For these variable sources, the median difference between 2MASS and UKIDSS after 
subtracting the 2MASS error is 0.05, 0.06, 0.05\,mag for all objects and 0.07, 0.09, 0.06 for those 
with disks.

This analysis shows that a) about half of the YSOs in IC348 exhibits low-level variability in near-infrared
bands, typically on the level of a few percent, and b) the level of variations is slightly enhanced in 
objects with disks.

\subsection{NGC1333}

NGC1333 is an extremely young cluster with an age of $\sim 1$\,Myr and a distance of $\sim 300$\,pc. As 
IC348, NGC1333 is located in the Perseus star forming region. We use the sample of 137 Class I and II sources 
identified by \citet{2008ApJ...674..336G} based on Spitzer data, which should be essentially free from 
contamination. As these objects are identified from the infrared excess, all are considered to have a 
disk (or at least circumstellar material).

The cluster has been observed by 2MASS from November 1999 to October 2000 and by UKIDSS/GPS early 2007, i.e.
the average epoch difference is $\sim 7$\,yr.

\begin{figure*}
\includegraphics[width=6.1cm,angle=-90]{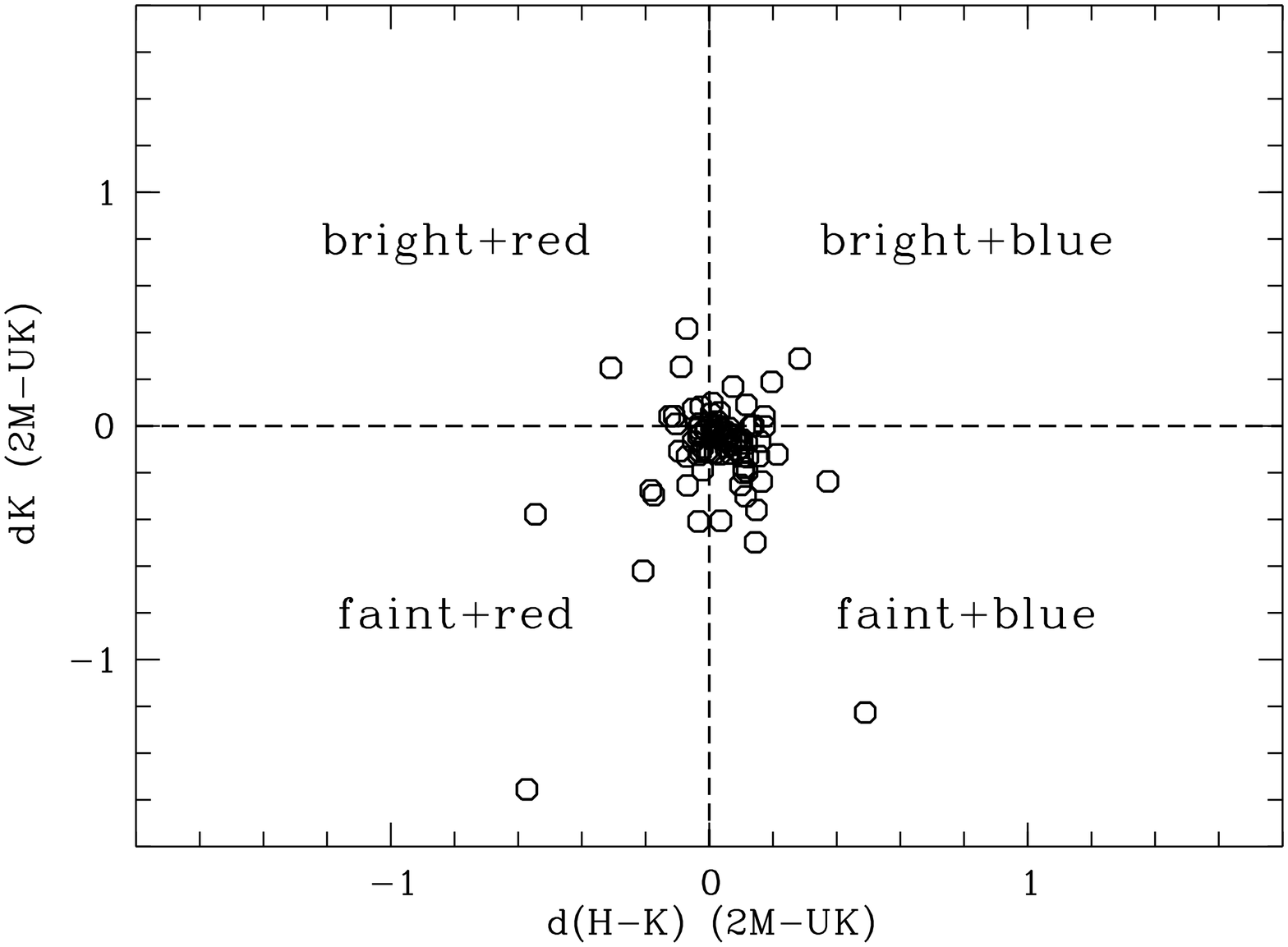} \hfill
\includegraphics[width=6.1cm,angle=-90]{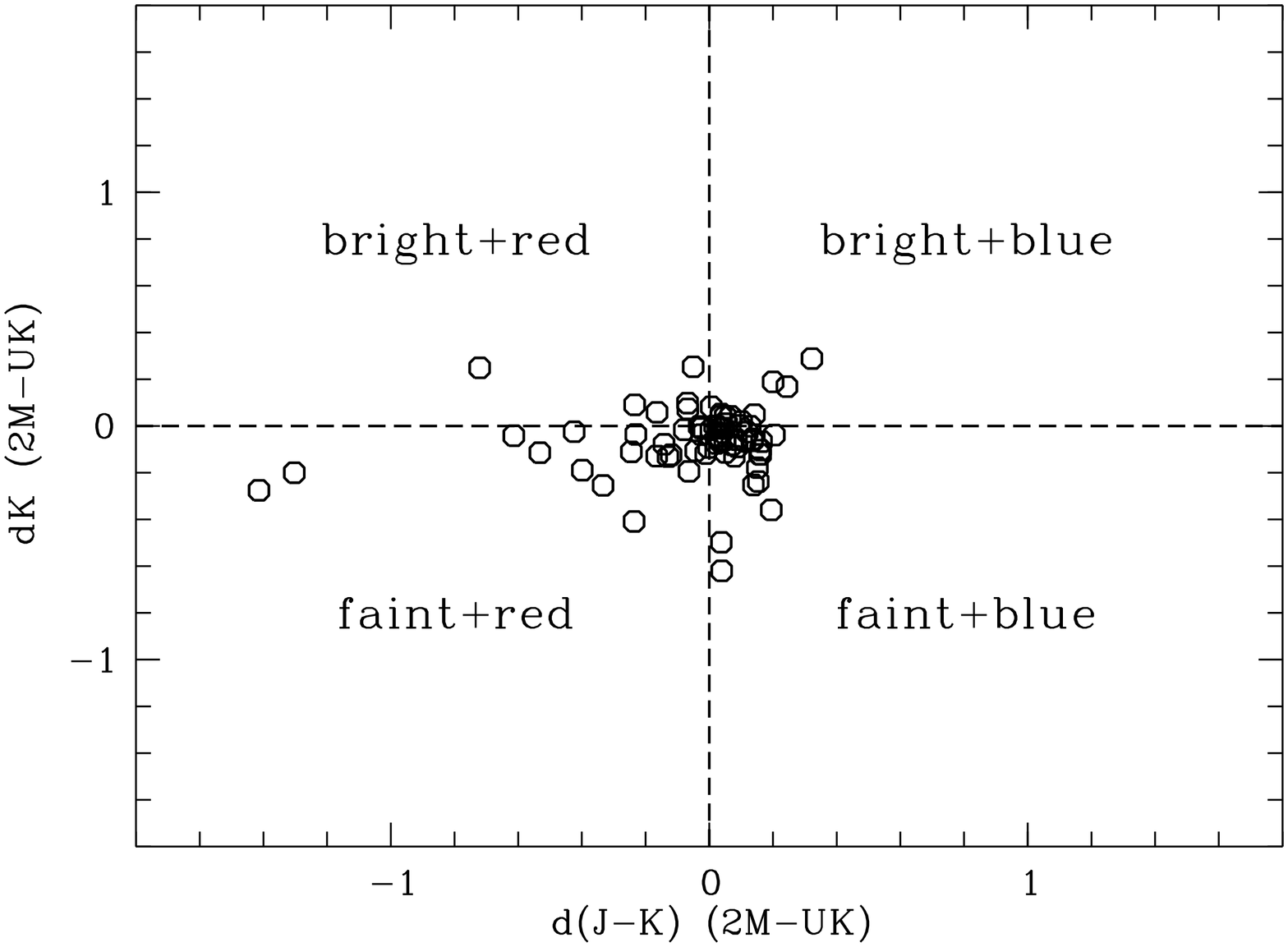} 
\caption{K-band variability vs. colour variability in $H-K$ (left panel) and $J-K$ (right panel) for YSOs in NGC1333
on timescales of $\sim 7$\,yr. The plots use data from 2MASS and UKIDSS.
\label{f6}}
\end{figure*}

From the initial sample, 96 have K- and H-band photometry in the two surveys, 82 of them unaffected
by saturation. 84 have K- and J-band photometry, 72 unaffected by saturation. In Fig. \ref{f6} the
usual plots are shown, for the $H-K$ colour (right panel) and the $J-K$ colour (left panel). 

Three objects appear in the plots at $dK>0.5$, one of them in both panels. All three have $>0.5$\,mag amplitudes
in one more band.  These three are contained in Table 
\ref{var}. As indicated, all of them are considered to have disks. Compared with the other regions, the
upper limit of the amplitudes is somewhat higher in NGC1333 (1.6 in K, instead of 1.0), possibly due to the
extreme youth and early evolutionary stage of the cluster. The frequency of highly variable objects is
3/82 or 4\%. 

Some more objects have large variations in the J-band up to $dJ>1$, but are much less variable in the other bands. 
It turns out that they are close to the sensitivity limit of 2MASS in the J-band and in some cases affected by 
neighbouring brighter stars. Although they need verification with additional epochs, at this point they are not
considered to be variable.

Similar to IC348, I determine the level of variability in the entire sample. The median difference in the three 
bands J, H, and K is 0.09, 0.09, 0.09\,mag. The fraction of objects with 2$\sigma$ variability is 47, 40, and 60\%. 
These fractions have typical binomial 1$\sigma$ errors of 6\%. For the variable objects, the median difference
between 2MASS and UKIDSS after subtracting 2MASS errors is 0.20, 0.14, 0.11\,mag in the three bands J, H, K.
As noted above, all objects analysed in NGC1333 show evidence for the presence of a disk. The fraction
of variable objects is similar to IC348, but the typical level of variations is higher in NGC1333 by a factor
of $\sim 2$, maybe because the cluster is younger than IC348. 

For comparison, variable stars in the ONC, a cluster similar in age to NGC1333, show median peak-to-peak amplitudes 
of 0.15, 0.12, and 0.11\,mag in the three bands J, H, and K on timescales of about one month \citep{2001AJ....121.3160C}. 
This is consistent with the values in NGC1333, indicating that the low-level variability mostly occurs on timescales 
of days to weeks, not years.

\subsection{$\sigma$\,Ori}

For the open cluster $\sigma$\,Ori, age $\sim 3$\,Myr, \citet{2009A&A...505.1115L} recently
published a comparison between 2MASS and UKIDSS data. They look at 263 sources with J-band
magnitudes between 12 and 16. Two of them show J-band variability by $dJ \sim 1.0$, three more with 
$dJ>0.5$. Thus the frequency of highly variable objects is in the range of 2\% for the total sample or, assuming 
a disk fraction of 30\%, 6\% for the objects with disks. These numbers are derived for the J-band, whereas the
K-band is used in the other regions, but they confirm that highly variable objects are rare (well below 10\%)
in star forming regions.

\section{Discussion}
\label{s3}

\subsection{Fraction of highly variable objects}

The main result from the analysis in Sect. \ref{s2} is that strongly variable objects on timescales of several
years are rare among Class II YSOs. Adding $\rho$-Oph, ONC, IC348, and NGC1333, there are 11 sources with $>0.5$\,mag
variations in two near-infrared bands on timescales 6-8\,yr, out of $\sim 620$ YSOs, from which $\sim 320$
have disks (here I assume a disk frequency of 50\% for the ONC). The
fraction of such variable objects is thus 2\% for the full sample and 3\% for YSOs with disks (Class I or
Class II). These low fractions are confirmed by the independent study in $\sigma$\,Ori 
\citep{2009A&A...505.1115L}. They are much lower than the number reported in 
\citet[][20-25\%]{2009MNRAS.398..873S}. This may be due to the much smaller sample size in the previous
study (only about 20 accretors) or due to an overestimate in the assumed contamination for their sample of
photometrically selected YSO candidates. In the current study, only high-confidence members of star forming
regions are considered, which should provide a more reliable result.

These results have implications for the interpretation of the spread in HR diagrams that is ubiquitously 
observed in star forming regions. Variability in the optical and near-infrared bands will to some extent contribute 
to this spread, as the luminosities are routinely estimated from I- or J-band photometry which is considered 
to give the best estimate of the photospheric flux. In luminosity, the spread in HR diagrams for very young 
regions is often as much as 1 order of magnitude \citep[e.g.][]{2003ApJ...593.1093L}. 

The results in this study indicate that the variability in the J-band is less than 0.5\,mag for the overwhelming 
majority of YSOs, i.e. a factor of $<1.6$ in luminosity, which is only a minor part of the observed spread in the 
HR diagrams. Moreover, the {\it typical} level of variability in the J-band is much lower than that. About 
half the objects show variations of 5-20\%, depending on the disk fraction and possibly age, the other half is
not variable within a few percent, consistent with previous results by \citet{2001AJ....121.3160C}.
Variability on timescales up to 6-8\,yr can definitely cause outliers in HR diagrams and has to be taken into 
account when deriving fundamental parameters for individual sources, as argued in \citet{2009MNRAS.398..873S}. 
For the analysis of the stellar properties of large samples, however, it is insignificant. This finding is 
similar to the conclusions drawn by \citet{2001AJ....121.3160C} and \citet{2005MNRAS.363.1389B}.

\subsection{Nature of highly variable objects}

As far as infrared data is available, 10/11 variable sources show evidence for the presence of a circumstellar 
disk. Since the disk fractions in the samples are high, this does not imply a statistically significant 
connection between variability and disk, but it gives several options to explain the nature of the variability.
Without disks, only two options remain, eclipses by a companion and chromospheric flares. Both are relatively
short events on timescales of hours and thus unlikely to be detected with only two epochs. 
With an accretion disk, additional explanations become viable: a) variations in the hot spots generated
by an accretion shock, b) variable circumstellar extinction, c) variable emission from a dusty disk, d)
eclipses by optically thick circumstellar material \citep{2001AJ....121.3160C}.

With two epochs it is not possible to unambiguously decide between these options, however, the colour
of the variability gives a first hint. Hot spots and extinction cause decreasing amplitudes towards longer
wavelengths, i.e. $dK < dJ$ or $dK< dH$. The reverse is the case for variable disk emission. Eclipses
would cause grey absorption, i.e. similar amplitudes in different bands. As seen in Table \ref{var},
5/13 clearly fall in the first category, 2/13 in the second, and 6/13 in the third.

With reddening laws for standard dust properties, variable extinction causes J-band amplitudes that are 
1.5 times larger than in H-band and 2.5 times larger than in the K-band \citep{2009MNRAS.398..873S}.
3 objects from Table \ref{var} fit these requirements (ISO-Oph-21, ISO-Oph-65, MBO46). The other two
with decreasing amplitude towards longer wavelength are better explained by variable hot spots, which
causes less colour variations than extinction. 

For a definite decision about the nature of the variability, follow-up spectroscopy and further multi-band
monitoring is required. As noted in Sect. \ref{s2}, the current sample might still be affected by
crowding, companions or other problems with the photometry. Particularly the 6 objects with almost equal
amplitudes in several bands seem highly interesting for follow-up, as they might contain equivalents to
KH15D \citep{2002PASP..114.1167H} which could give detailed insights into the structure of the disk.

\subsection{Episodic accretion}

One other possible application for the results presented in this paper is to derive constraints for 
scenarios of 'episodic accretion', i.e. the hypothesis that a large fraction of the stellar mass 
is accreted in short episodic bursts of accretion. Episodic accretion might be the explanation for the 
dramatic outbursts in FU Ori-type objects, and it could also resolve the problem of the underluminosity of 
protostars \citep[e.g.][]{1996ARA&A..34..207H}. Based on Spitzer photometry, \citet{2009ApJS..181..321E} find 
that half the mass of T Tauri stars is accreted in only 7\% of the Class I lifetime, which is $\sim 0.5$\,Myr. 
This implies that episodes with strong accretion of $>10^{-5}\,M_{\odot}$yr$^{-1}$ which last in total 
a few $10^5$ ys are interspersed with significantly longer quiescent phases with much lower accretion 
rates ($<10^{-6}\,M_{\odot}$yr$^{-1}$). These numbers are supported by the available submm data from
embedded protostars \citep{2009ApJ...692..973E}. 

There is strong interest in episodic accretion from theoretical work on star and planet formation.
Numerical simulations of the gravitational cloud collapse \citep{2006ApJ...650..956V,2009ApJ...704..715V} 
actually predict episodic accretion to occur, with duty cycles that are consistent with the current
constraints by the observations (e.g., 1-2\% of protostars with accretion rates 
$>10^{-5}\,M_{\odot}$yr$^{-1}$). Models of episodic accretion can reproduce the observed luminosity 
spread in HR diagrams, with quiescent phases of $10^3$ to $10^4$\,yr \citep{2009ApJ...702L..27B}.
These 'lulls' allow for the formation of low-mass stars, brown dwarfs, and planets via disk 
fragmentation, and their duration may be critical for the frequency of these objects 
\citep{2011ApJ...730...32S}.

Episodic accretion causes variability on very long timescales of hundreds of years or
more. One way of improving the constraint on the aforementioned models is thus to monitor the brightness of large 
samples of accreting YSOs over long time windows, to identify possible outbursts and derive their
frequency. The increase in accretion rate should be approximately proportional to the increase in
bolometric luminosity, if the gravitational energy of the accreted material is fully converted to radiation.

It is difficult to assess the effect of an accretion burst on the near-infrared magnitudes and colours,
without knowing the spectral energy distribution of the accretion shockfront and the effects of increased
accretion on the heating and emission of the disk. In contrast to most previously detected accretion bursts, 
two recent events have been observed in the near-infrared and indicate $dK$ of 3-4\,mag and and $dJ$ of 3-5\,mag
\citep{2011A&A...526L...1C,2011ApJ...730...80M}. Although a full characterisation is still pending, these 
two events probably do not classify as FU Ori-type outbursts \citep[see][]{2011A&A...527A.133K}; they are 
more likely to be weaker bursts with accretion rate increase of at most 2-3 orders of magnitude. 

In the samples investigated in this paper, there are zero objects with K- or J-band magnitude increase
by more than 2\,mag, i.e. it is unlikely that any of them undergoes an accretion rate outburst by 
several orders of magnitude, as it would be required in the episodic accretion scenario. This sets
a lower limit on the duty cycle of accretion bursts in YSOs. Since this study covers about 320 objects 
with accretion disks and timescales of 6-8\,yr, the quiescent phases with low accretion rates last
at least 2000-2500\,yr. We note that \citet{2001AJ....121.3160C} determined a lower limit of 5400\,yr 
for the duty cycle of FU Ori outbursts, using a similar approach. Their sample, however, comprises the
entire YSO population of Orion A, including objects without disks. Assuming that about half of their
objects have disks, that timescale drops to 2700\,yr, very similar to our result.

While this is consistent with the constraints from the infrared luminosities of protostars (see above),
it is important to point out that the majority of the objects considered here are Class II sources, 
because the earlier stages of the protostellar evolution (Class 0 and I) have shorter lifetimes and
are heavily embedded and thus often not observable in the near-infrared. It is possible that accretion bursts
become less frequent as the objects progress towards the Class II stage. Maybe we should expect much longer
duty cycles (or possibly weaker bursts) at Class II compared with Class I stage. 

Thus, the constraint given above is currently of limited use when comparing with the model predictions 
for the earlier stages. Improving the analysis requires one of the following: 1) Adding second 
epoch data for a number of southern star forming regions with well-characterised YSO population not 
observed by UKIDSS (e.g., Chamaeleon, Lupus, Corona Australis), to increase the sample size substantially. 
2) Establishing a census of YSOs in star forming region for which several epochs are already available 
(e.g., from Galactic plane surveys). 3) Carrying out a similar analysis in the mid-infrared regime where 
the Class I sources are visible (e.g., based on data from Spitzer or WISE, see for example \citet{2011ApJ...733...50M}). 
4) Extending the time baseline by adding more epochs over the next decades, which becomes feasible through all-sky 
monitoring facilities like Pan-STARRS or LSST. 

\section*{Acknowledgments}
I would like to thank Dawn Peterson, Caroline d'Angelo, Duy Cuong Nguyen, Gwendolyn Meeus, Dirk Froebrich, and
Alexis Brandeker for valuable advice during the writing of this paper. The referee report from Keivan Stassun
has helped to improve the paper. A student project carried out by Zoe Lematy from Domician College Wicklow 
helped to initiate this project. Part of this work was funded by the Science Foundation Ireland through grant no. 
10/RFP/AST2780.
 
\newcommand\aj{AJ} 
\newcommand\actaa{AcA} 
\newcommand\araa{ARA\&A} 
\newcommand\apj{ApJ} 
\newcommand\apjl{ApJ} 
\newcommand\apjs{ApJS} 
\newcommand\aap{A\&A} 
\newcommand\aapr{A\&A~Rev.} 
\newcommand\aaps{A\&AS} 
\newcommand\mnras{MNRAS} 
\newcommand\pasa{PASA} 
\newcommand\pasp{PASP} 
\newcommand\pasj{PASJ} 
\newcommand\solphys{Sol.~Phys.} 
\newcommand\nat{Nature} 
\newcommand\bain{Bulletin of the Astronomical Institutes of the Netherlands}
\newcommand\memsai{Mem. Soc. Astron. Ital.}

\bibliographystyle{mn2e}
\bibliography{aleksbib}

\label{lastpage}

\end{document}